\documentclass[pageno]{jpaper}

\usepackage[normalem]{ulem}
\usepackage{afterpage}
\usepackage{amsfonts}
\usepackage{amsmath}
\usepackage{amsthm}
\usepackage{balance}
\usepackage{booktabs}
\usepackage{clrscode3e}
\usepackage{hhline}
\usepackage{listings}
\usepackage{mathtools}
\usepackage{microtype}
\usepackage{multirow}
\usepackage{placeins}
\usepackage{stmaryrd}
\usepackage[super]{nth}
\usepackage{url}
\usepackage{xargs}
\usepackage{xspace}
\usepackage{todonotes}
\usepackage{textcomp}
\usepackage{xcolor}
\usepackage{wrapfig}
\usepackage{enumitem}
\usepackage[toc,page]{appendix}
\usepackage{etoolbox}

\definecolor{varcolor}{RGB}{46, 139, 87}

\let\origthelstnumber\thelstnumber
\makeatletter
\newcommand*\Suppressnumber{%
  \lst@AddToHook{OnNewLine}{%
    \let\thelstnumber\relax%
     \advance\c@lstnumber-\@ne\relax%
    }%
}

\newcommand*\Reactivatenumber{%
  \lst@AddToHook{OnNewLine}{%
   \let\thelstnumber\origthelstnumber%
   \advance\c@lstnumber\@ne\relax}%
}

\lstdefinestyle{makefile}{
  moredelim=[is][\bfseries\color{blue}]{\{:}{:\}},
  moredelim=[s][\color{varcolor}]{\$(}{)}
}

\lstdefinestyle{rikerfile}{
  language=bash,
  numbers=left,
  xleftmargin=6pt,
  numbersep=4pt,
  numberstyle=\tiny\color{darkgray},
  basicstyle=\ttfamily\footnotesize\color{black}
}

\renewcommand{\paragraph}[1]{\subsubsection*{#1}}

\SetTracking{encoding={*}, shape=sc}{0}
\lstdefinelanguage{TraceIR}{
  keywords={Launch, PathRef, ExpectResult, MatchMetadata, MatchContent, UpdateContent, FileRef, DirRef, PipeRef, SymlinkRef, SpecialRef},
  basicstyle=\ttfamily\footnotesize\color{black},
  rulecolor=\color{black},
  keywordstyle=\color{black}\bfseries,
  ndkeywords={gid, type, perms, mtime, hash, cached, uid},
  ndkeywordstyle=\color{darkgray}\bfseries,
  identifierstyle=\color{blue},
  sensitive=false,
  comment=[l]{//},
  morecomment=[s]{/*}{*/},
  commentstyle=\color{black}\ttfamily,
  stringstyle=\color{blue}\ttfamily,
  morestring=[b]',
  morestring=[b]"
}

\lstdefinelanguage{TraceIRSmall}{
  keywords={Launch, PathRef, ExpectResult, MatchMetadata, MatchContent, UpdateMetadata, UpdateContent, FileRef, DirRef, PipeRef, SymlinkRef, SpecialRef, ExitResult, AddEntry, RemoveEntry, UsingRef, DoneWithRef, Exit, CompareRefs},
  basicstyle=\ttfamily\scriptsize\color{black},
  rulecolor=\color{black},
  keywordstyle=\color{black}\bfseries,
  ndkeywords={gid, type, perms, mtime, hash, cached, uid},
  ndkeywordstyle=\color{darkgray}\bfseries,
  identifierstyle=\color{blue},
  sensitive=false,
  comment=[l]{//},
  morecomment=[s]{/*}{*/},
  commentstyle=\color{black}\ttfamily,
  stringstyle=\color{blue}\ttfamily,
  morestring=[b]',
  morestring=[b]",
  numbers=none,
  xleftmargin=0cm,
  escapechar=|
}

\newtoggle{arxiv}
\toggletrue{arxiv}

\iftoggle{arxiv}{
  \newcommand{\riker}{\textsc{LaForge}\xspace}
}{
  \newcommand{\riker}{\textsc{Riker}\xspace}
}
\newcommand{\shake}{\textsc{Shake}\xspace}
\newcommand{\fabricate}{\textsc{Fabricate}\xspace}
\newcommand{\memoize}{\textsc{Memoize}\xspace}

\newcommand{\tup}{\textsc{Tup}\xspace}
\newcommand{\pluto}{\textsc{Pluto}\xspace}
\newcommand{\buck}{\textsc{Buck}\xspace}
\newcommand{\bazel}{\textsc{Bazel}\xspace}
\newcommand{\rattle}{\textsc{Rattle}\xspace}
\newcommand{\vesta}{\textsc{Vesta}\xspace}
\newcommand{\ninja}{\textsc{Ninja}\xspace}
\newcommand{\cloudbuild}{\textsc{CloudBuild}\xspace}
\newcommand{\cereal}{\textsc{Cereal}\xspace}
\newcommand{\rr}{\textsc{RR}\xspace}
\newcommand{\cliEleven}{\textsc{CLI11}\xspace}
\newcommand{\blake}{\textsc{BLAKE3}\xspace}

\iftoggle{arxiv}{
  \newcommand{\rkr}{\texttt{laf}\xspace}
}{
  \newcommand{\rkr}{\texttt{rkr}\xspace}
}
\iftoggle{arxiv}{
  \newcommand{\rkrdir}{\texttt{.laf}}
}{
  \newcommand{\rkrdir}{\texttt{.rkr}}
}

\newcommand{\make}{\texttt{make}\xspace}
\newcommand{\cmake}{\texttt{cmake}\xspace}
\newcommand{\gcc}{\texttt{gcc}\xspace}
\newcommand{\clang}{\texttt{clang}\xspace}
\newcommand{\Makefile}{\texttt{Makefile}\xspace}
\iftoggle{arxiv}{
  \newcommand{\rikerfile}{\texttt{Buildspec}\xspace}
}{
  \newcommand{\rikerfile}{\texttt{Rikerfile}\xspace}
}
\newcommand{\ptrace}{\texttt{ptrace}\xspace}
\newcommand{\seccomp}{\texttt{seccomp}\xspace}
\newcommand{\traceir}{TraceIR\xspace}

\newcommand{\EvalStmt}[5]{\proc{EvalStmt}(#1, #2, #3, #4, #5)\xspace}
\newcommand{\CmdOf}[1]{\proc{CmdOf}(#1)\xspace}

\newcommand{\IRCmdRef}{\proc{CmdRef}\xspace}
\newcommand{\IRRef}{\proc{Ref}\xspace}
\newcommand{\IRRefComparison}{\proc{RefComparison}\xspace}

\newcommand{\IRBool}{\proc{Bool}\xspace}
\newcommand{\IRInt}{\proc{Int}\xspace}
\newcommand{\IRString}{\proc{String}\xspace}
\newcommand{\IRAccessFlags}{\proc{AccessFlags}\xspace}
\newcommand{\IRMetadataState}{\proc{MetadataState}\xspace}
\newcommand{\IRContentState}{\proc{ContentState}\xspace}
\newcommand{\IRCommand}{\proc{Command}\xspace}

\newcommand{\IRLaunch}{\proc{Launch}\xspace}

\newcommand{\IRExpectResult}{\proc{ExpectResult}\xspace}
\newcommand{\IRJoin}{\proc{Join}\xspace}

\newcommand{\IRMatchMetadata}{\proc{MatchMetadata}\xspace}

\newcommand{\IRPathRef}{\proc{PathRef}\xspace}

\newcommand{\numbench}{14\xspace}
\newcommand{\percentslowdown}{16.1\%\xspace}
\newcommand{\absslowdown}{3.08s\xspace}
\newcommand{\benchA}{LLVM\xspace}
\newcommand{\benchB}{memcached\xspace}

\newcommand{\punt}[1]{}
\newcommand{\sectref}[1]{\S\ref{#1}}

\newcommand{\cmd}[1]{\texttt{#1}}

\newcommand{\fun}[1]{\texttt{#1}}
\newcommand{\file}[1]{\texttt{#1}}

\newcommand{\wrapperOverheadReduction}{10\%\xspace}

\author{
    Charlie Curtsinger\\
    Grinnell College\\
    curtsinger@grinnell.edu
  \and
    Daniel W. Barowy\\
    Williams College\\
    dbarowy@cs.williams.edu
}

\begin{document}

%% Paper name
\iftoggle{arxiv}{
  \title{\riker: Always-Correct and Fast Incremental Builds from Simple Specifications}
}{
  \title{\riker: Don't \make, Make it So}
}

\date{}
\maketitle

\thispagestyle{empty}

\begin{abstract}

Developers rely on build systems to generate software from code.
At a minimum, a build system should produce build targets from a clean copy of the code.
However, developers rarely work from clean checkouts.
Instead, they rebuild software repeatedly, sometimes hundreds of times a day.
To keep rebuilds fast, build systems run \emph{incrementally}, executing commands only when built state cannot be reused.
Existing tools like \make present users with a tradeoff.
Simple build specifications are easy to write, but limit incremental work.
More complex build specifications produce faster incremental builds, but writing them is labor-intensive and error-prone.
This work shows that no such tradeoff is necessary;
build specifications can be both simple and fast.

We introduce \riker, a novel build tool that eliminates the need to specify dependencies or incremental build steps.
\riker builds are easy to specify;
developers write a simple script that runs a full build.
Even a single command like \texttt{gcc src/*.c} will suffice.
\riker traces the execution of the build and generates a transcript in the \traceir language.
On later builds, \riker evaluates the \traceir transcript to detect changes and perform an efficient incremental rebuild that automatically captures all build dependencies.

We evaluate \riker by building \numbench software packages, including \benchA and \benchB.
Our results show that \riker automatically generates efficient builds from simple build specifications.
Full builds with \riker have a median overhead of \percentslowdown compared to a project's default full build.
\riker's incremental builds consistently run fewer commands, and most take less than \absslowdown longer than manually-specified incremental builds.
Finally, \riker is always correct.

% Alternative phrasing:
%\riker's overhead on full builds is just \percentslowdown on average, and incremental builds with \riker consistently run fewer commands than a project's default build system.
%Most incremental builds with \riker complete within \absslowdown of a manually-specified incremental build, despite the much simpler build specification.

\end{abstract}

\section{Introduction}
\label{sec:intro}

% Build systems are important
% Builds should be correct and fast
An important but often under-appreciated component of software is its \emph{build system}.
Build systems specify how code and other assets should be transformed into executable software.
They capture compilation procedures left unstated in the source code itself. %, such as how and in what order software components must be built.
Critically, build systems make the process of building software more reliable since the programmer need not remember and correctly reproduce the sequence of steps necessary to produce working executables.
Build systems should satisfy two sometimes-competing goals: builds must be correct, and they must be fast.

% Introduce monolithic builds with make
To illustrate the challenge in making builds both fast and correct we begin with an example build system that uses \make, one of the earliest and most widely-used build tools~\cite{doi:10.1002/spe.4380090402}.
A \make-based build is specified in a domain-specific language, and stored in a file called a \Makefile.
The \make tool is responsible for reading this specification and performing the build.
The simplest build to specify with \make is one that performs a \emph{monolithic build} like the one below, which builds a program from three source files and two headers.

\begin{lstlisting}[style=makefile]
{:program:}: main.c x.c x.h y.c y.h
  gcc -o program main.c x.c y.c
\end{lstlisting}

% Monolithic builds are correct, but not fast.
% Introduce incremental builds with make
It is easy to see that this build is correct, because all dependencies are specified and only a single build command is needed.
Unfortunately, a monolithic build specification will rerun the entire build process even when only a subset of build dependencies have changed.
For example, changing \file{x.c} will cause \gcc to recompile all three \file{.c} files, even though \file{y.c} and \file{main.c} are unchanged.

The cost of full builds is low for small projects, but building large projects can take much longer.
For example, a full build of LLVM takes nearly 20 minutes on a typical developer workstation;
this is far too long for a developer to wait to test a small code change.
To address this, many large projects use build systems that perform \emph{incremental rebuilds}.
An incremental rebuild runs only the set of commands whose inputs have changed since the last build, resulting in dramatic speedups.
To specify an incremental build for \make, developers break the build into fine-grained steps and list the dependencies for each step.
The following \Makefile specifies an incremental build for the same example program:

\begin{lstlisting}[style=makefile]
{:program:}: main.o x.o y.o
  gcc -o program main.o x.o y.o
{:main.o:}: main.c x.h y.h
  gcc -c -o main.o main.c
{:x.o:}: x.c
  gcc -c -o x.o x.c
{:y.o:}: y.c
  gcc -c -o y.o y.c
\end{lstlisting}

% Incremental builds can be faster, but correctness is an issue.
% Auto-generated dependency information is not sufficient.
This alternative specification states how to build each intermediate \file{.o} file from its source files, and how those \file{.o} files are combined into the final output executable \texttt{program}.
Now, modifying \file{x.c} no longer triggers a full rebuild.
Instead, the build only generates a new \file{x.o} which \gcc then links with the other \file{.o} files already on disk.
This version provides a clear performance improvement, taking advantage of the fact that developers rarely modify all files between rebuilds.

This build specification also illustrates the dangers inherent in a complex build specification: missing dependencies.
Suppose \file{x.c} includes \file{x.h};
with the above \Makefile, changing \file{x.h} will not trigger a rebuild of \file{x.o} as it should.
A developer with a previously-built working copy could end up building a different executable than another developer with a clean copy of exactly the same source code.
A key correctness property of an incremental build system is that it should always produce the same output as a full build.
Incorrect builds waste developer time and can introduce latent errors in released software.
% A senior developer at a startup described a build system error that caused full builds to fail while incremental builds on developer machines succeeded, wasting several hours of developer time.
% These errors occur with alarming frequency;
Such errors are endemic to build specifications;
a recent study showed that more than two-thirds of the open-source programs analyzed had serious build specification errors~\cite{10.1145/3428212}.

% What about workarounds for capturing dependency information?
%To mitigate the risk of missed dependencies, developers sometimes over-specify dependencies.
%For example, the \Makefile above could be modified to list every \file{.h} file as a dependency for each \file{.o} target.
%Although over-specification ensures a correct build, it will also make rebuilds less incremental.
To mitigate the risk of missed dependencies, developers sometimes use \gcc's dependency generation feature (\texttt{-MMD} and \texttt{-MP} flags) which produces a list of dependencies that can be included in a \Makefile.
Dependency generation is only available for compilers with explicit \make support, and still it requires that users manually insert incremental build targets.
The feature also does not work well for projects that generate code or use multiple programming languages.
% and developers still need to make complex, manual changes to their \Makefile to use it.
% Even projects that use languages with integrated build systems like Go and Rust will miss dependencies in multi-language projects, or when projects use generated code or dynamic dependencies.
% These unsupported cases will still require developers to specify dependencies manually, and to keep those dependencies up to date as code evolves.

%\todo[inline]{CC: Should we discuss problems with parallel make in the intro? It's not totally clear where it fits in, but doing it correctly is a big contribution of this work. DWB: Definitely need to mention this as early as possible.  It would be great to have a simple example that breaks make -j.}

% Introduce our work
% Existing build tools force users to choose between a simple, correct build system, or a complex, efficient build system, but this does not have to be the case.
This paper introduces \riker, a \emph{forward build tool} that gives developers the benefits of an incremental build system with the simplicity of monolithic builds.
Forward build tools do not require users to specify any dependencies at all.
Instead, \riker uses system call tracing to precisely identify dependencies for all commands, including subcommands.
\riker produces efficient incremental rebuilds even when users write monolithic build specifications.
For example, many projects can be built using a single build command such as \texttt{gcc *.c}, and \riker discovers incremental builds from even this simple specification.
\riker captures dependencies for the C compiler (\cmd{cc1}), assembler (\cmd{as}), and linker (\cmd{ld} and \cmd{collect2}), even though the user only invoked the \gcc driver program.
%On rebuild, \riker only runs the set of commands strictly necessary to update the build.
%\riker automatically updates dependency information, ensuring build correctness.
Put simply, \riker allows users write build specifications that are both simple and efficient.
% Finally, migrating to \riker from an existing build system requires near-zero effort.
% Users can simply instruct \riker to run their existing build, and it will automatically discover dependencies, running future builds incrementally and in parallel.

% Contributions
\paragraph{Contributions.} This paper makes the following contributions:

\begin{itemize}
   \item We introduce \traceir, an intermediate representation that captures the effects and dependencies of build commands.
    \traceir encodes interactions with paths, files, directories, pipes, and more;
    it enables correct handling of circular and temporal dependencies, which both occur in real builds.
  \item We introduce the \riker algorithm, which generates and evaluates \traceir to run correct and fast incremental builds without manual specification.
    Ours is the first forward build algorithm that can run efficient incremental builds without manually-specified incremental build steps.
  \item Finally, we present an implementation of \riker for Linux, which we evaluate by building \numbench real-world software projects including \benchA and \benchB.
\end{itemize}

\section{Overview}
\label{sec:overview}

\riker builds software using a simple specification called a \rikerfile.
A \rikerfile is typically a short shell script that runs a full build, although it can be any executable that performs the build.
On the first build, \riker runs the \rikerfile under a lightweight form of system call tracing~\cite{10.5555/231070}.
As the \rikerfile executes, \riker generates a \emph{transcript} of the build in the \traceir language.
\traceir is a \emph{program} that describes a build's sequence of operations and their outcomes.
%For example, a build might resolve a path, read a file, or launch a sub-command.
%Outcomes could be that a path resolves with an \texttt{EPERM} error, or that a read observes a file with specific contents or metadata.

When the user requests a rebuild, \riker evaluates the stored \traceir program instead of running the full build.
Evaluating \traceir updates an in-memory \emph{model} of the filesystem.
Any statement in the \traceir program that returns an expected outcome doubles as a \emph{predicate};
if the outcome changes, \riker knows that the command that produced the \traceir statement must run to update the build.
This phase of the build is performed entirely in memory, so checking is fast.

When \riker finds at least one command that must run, an incremental build is performed by re-evaluating the \traceir program, this time with a mix of emulation and actual command execution.
\riker emulates \traceir using the in-memory model for commands that do not need to run, effectively \emph{skipping} them.
\riker executes all other commands with system call tracing to generate new \traceir transcripts for those commands.
%Whenever a skipped command's outputs are visible to an executed command, \riker restores those outputs from its file cache.
%Whenever an executed command's outputs are visible to an emulated command, \riker observes the change and marks the emulated command to run.
\riker repeatedly re-evaluates the entire trace until no commands detect a change.
Under the vast majority of circumstances, commands are executed only once during a build (see~\sectref{sec:exitcodes}).
We describe this process in~\sectref{sec:building}.

\subsection{Examples}
\label{sec:examples}

% Start example riker build
We return to the working example from the introduction, a C language project with three source files and two headers.
The following \rikerfile suffices to build the program with \riker:

\begin{lstlisting}[language=bash]
#!/bin/sh
gcc -o program *.c
\end{lstlisting}

\paragraph{Running the first build.}
The \rkr shell command runs the \riker build program.
When \rkr is invoked with no saved state, \riker starts a full build, with tracing enabled, by executing the \rikerfile.

\begin{figure}[t]
  \begin{center}
    % anonymized figure
    \iftoggle{arxiv}{
	  \includegraphics[width=0.95\linewidth]{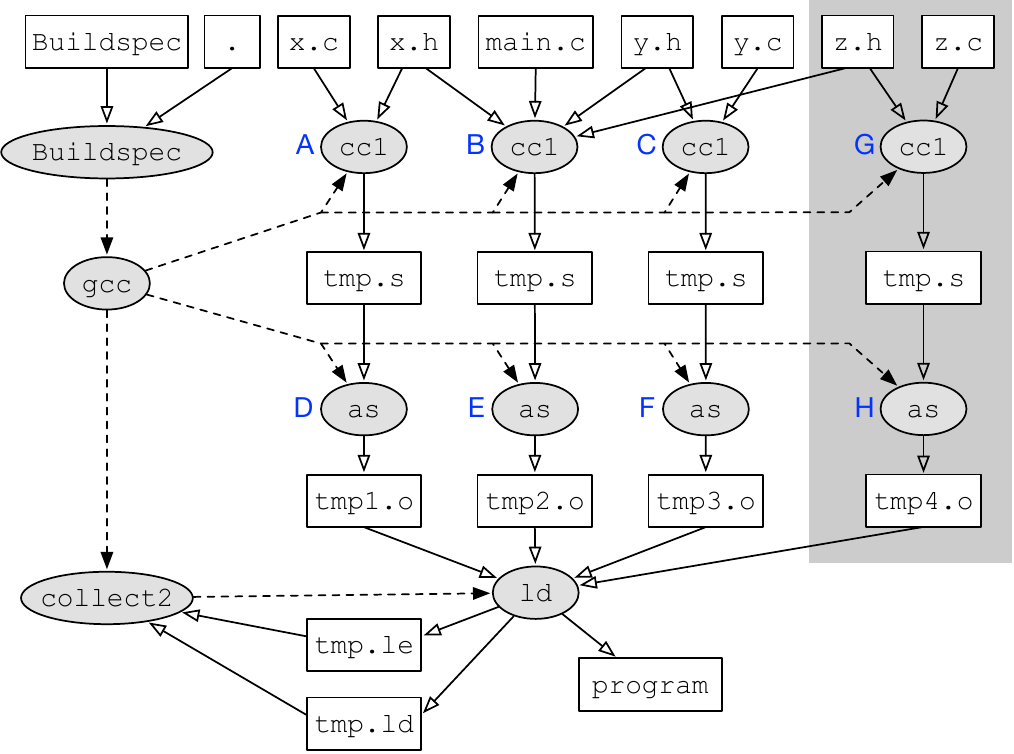}
	}{
	  \includegraphics[width=0.95\linewidth]{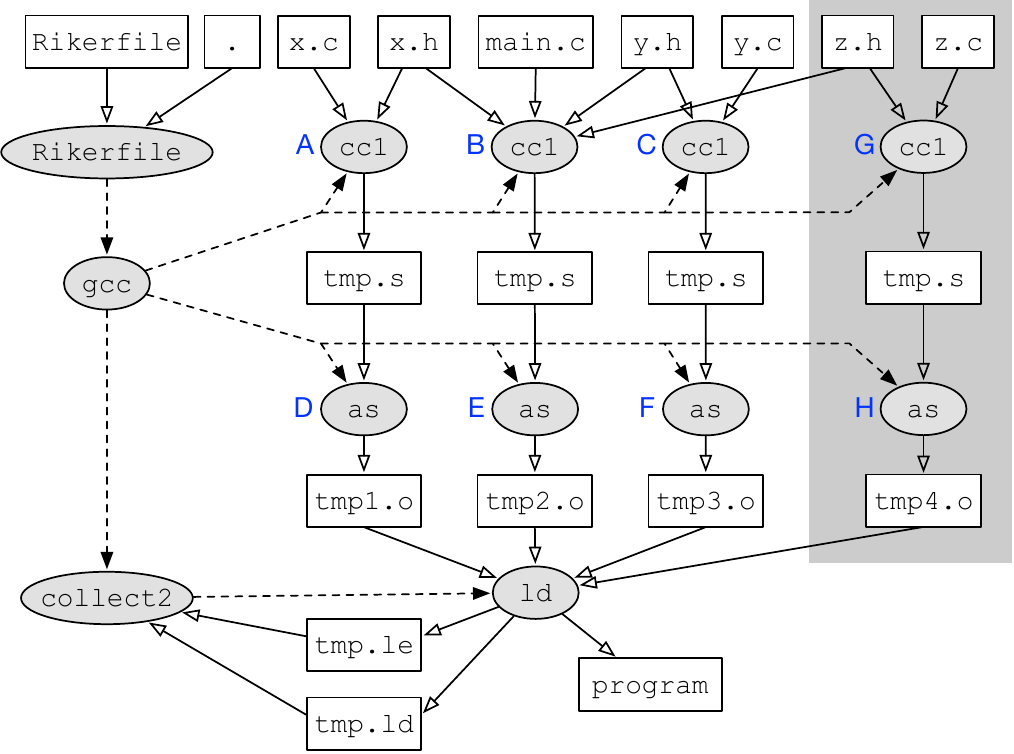}
	}
  \end{center}
  \vspace{-1.4em}
  \includegraphics[width=0.9\linewidth]{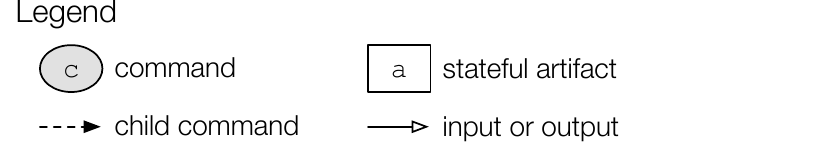}
  \caption{
    \small
    A dependency graph for the running example.
    Dashed edges show commands launching child commands, and solid edges indicate inputs and outputs.
    The grey box contains the modification induced by adding \file{z.h} and \file{z.c} to the build.
    \label{fig:example}
  }
\end{figure}

Figure~\ref{fig:example} shows that even simple builds have complex dependencies.
Oval vertices represent \emph{commands}, which correspond to programs run via \texttt{exec} system calls.
Rectangular vertices represent stateful \emph{artifacts} such as files or directories.
Dashed edges indicate that a parent command \emph{launched} a child command.
Solid edges indicate a command's input or output.
Although \riker captures dependencies on system includes, shared libraries, and the executable files for each command, we omit these in our example for clarity.

%\riker uses \ptrace to trace system calls, but incorporates a \seccomp BPF filter to exclude system calls that do not manipulate  and a small injected library inspired by~\cite{203227} to reduce the overhead of system call tracing.

The \rikerfile command launches \cmd{gcc}, which in turn launches three instances of \cmd{cc1}.
Each \cmd{cc1} instance compiles a \file{.c} file (and any included \file{.h} files) to a \file{.s} assembly file.
\cmd{gcc} also launches three instances of \cmd{as} to produce \file{.o} object files from each \file{.s} input.
Finally, \cmd{gcc} launches the \cmd{collect2} command, which launches the linker \cmd{ld}.
\cmd{ld} redirects \file{stdout} and \file{stderr} to temporary files which trigger \cmd{collect2} to conditionally re-link.
Note the cycle between \cmd{collect2} and \cmd{ld};
this particular cycle will be present in every build that uses \gcc.
Also observe that \gcc repeatedly reuses the same \file{tmp.s} temporary file, truncating it at the start of each \cmd{cc1} execution.
File reuse and cyclic dependencies are common features in builds, particularly those that use \gcc.
%\riker correctly handles both file reuse and cyclic dependencies.

We show the build graph in Figure~\ref{fig:example} for illustrative purposes only.
\riker itself does not store build information in graph form.
Instead, \riker stores a \traceir program.
Every intercepted system call translates into a small set of \traceir statements.
The following is an excerpt of the \traceir generated during the full build of the example:

%Importantly, a \traceir transcript is not a log of system calls, it is a \emph{program}.
%Although this program is critical to \riker's operation, users never see or manipulate \traceir at all.
%The \rkr tool generates \traceir while performing a build, and runs \traceir to plan rebuilds.

%\riker plans and runs incremental builds by evaluating this stored \traceir program (\sectref{sec:building}).
%More precisely, executing \traceir maintains an in-memory \emph{model} of the filesystem.
%When a \traceir statements representing a dependency is evaluated---like when a program opens or reads a file---its outcome is predicted in the model and then compared against the same action in the actual filesystem.
%Any difference between the two constitutes a \emph{change} made by the user.
%\riker marks the command associated with the observed change to run in order to update the build.
%Representing builds with \traceir is what allows \riker to correctly handle circular dependencies, file reuse, and other complex situations that come up in real builds.
%This is a departure from prior work on forward build systems, which universally use build dependence graphs to detect changes and plan rebuilds (\sectref{sec:related}).

\begin{lstlisting}[language=TraceIR, numbers=left, xleftmargin=0.5cm, escapechar=|]
sh_0 = Launch(|\rkr|, "sh |\rikerfile{}|", [...])|\label{line:dot-launch}|
r16 = PathRef(sh_0, CWD, ".", r--)|\label{line:dot-Ref}|
ExpectResult(sh_0, r16, SUCCESS)|\label{line:dot-expect}|
MatchMetadata(sh_0, r16, |\label{line:dot-metadata}\Suppressnumber|
 [uid=100, gid=100, type=dir, perms=rwxrwxr-x]) |\Reactivatenumber|
MatchContent(sh_0, r16, [dir:|\label{line:dot-content}\Suppressnumber|
 {"|\rikerfile{}|", "main.c", "x.c", "x.h", "y.c", "y.h"}]) |\Reactivatenumber|
\end{lstlisting}

Line~\ref{line:dot-launch} of the \traceir program corresponds to \riker's launch of the \rikerfile by a command it names \texttt{sh\_0}.
The remaining \traceir steps correspond to the \rikerfile's evaluation of the shell \texttt{*} glob character, which lists the current directory.
\texttt{sh\_0} opens the current directory, generating line~\ref{line:dot-Ref} of the \traceir.
The reference \riker names \texttt{r16}, returned on line~\ref{line:dot-Ref}, is local to the command \texttt{sh\_0}, and is the \traceir analogue of a file descriptor.
Line~\ref{line:dot-expect} is the first predicate in the \traceir program.
This \texttt{ExpectResult} statement encodes that the given path reference successfully \emph{resolves} to a file~\cite{pathresolution}.
Depending on the outcomes of path resolution is essential for correctness, because UNIX path searching results in many accesses that \emph{normally} return \texttt{ENOENT}, behavior that must be preserved on subsequent builds.
Line~\ref{line:dot-metadata} is generated because the \rikerfile issues an \texttt{fstat} system call to check the metadata of the \texttt{.} directory.
Finally, the predicate on line~\ref{line:dot-content} is generated when \rikerfile lists the current directory.

In the next section, we examine how the above program guides a rebuild after a user makes a code change.
We defer discussion of \traceir semantics to~\sectref{sec:traceir}.

\paragraph{Example 1: Adding a file.}
% How the trace helps a rebuild
After adding the files \file{z.c} and \file{z.h} and modifies \file{main.c} to include \file{z.h}, we run \rkr again to update the build.
The grey box in Figure~\ref{fig:example} shows the effect of the change on the build's dependence graph.
A good incremental build should not rebuild files unrelated to a change.
Here, \file{tmp1.o} and \file{tmp3.o} do not need updating since they do not depend---even transitively---on any of the changes.
At the very least, \cmd{cc1} and \cmd{as} should be called to compile \file{main.c} and \file{z.c}, and \cmd{collect2} and \cmd{ld} should be called to link the output to our preexisting object files.

\riker performs an incremental rebuild of the example by evaluating the \traceir from the previous build.
We assume the user does not change ownership or permissions for the current directory, so lines~\ref{line:dot-launch}--\ref{line:dot-metadata} evaluate just as before.
However, line~\ref{line:dot-content}, which depends on directory contents, reports a change because the current directory contains the new files \file{z.c} and \file{z.h}.
\riker therefore reruns and traces the \rikerfile.
When the \rikerfile command is rerun, the command's transcript is replaced with newly generated \traceir.

Although rerunning the \rikerfile might seem to imply that the entire build will run again, this is not the case.
When \rikerfile launches \gcc, \riker lets the execution proceed (also under tracing) because \gcc{}'s arguments, which which now include \file{z.c}, also change.
However, \riker \emph{skips} the commands labeled \texttt{A}, \texttt{C}, \texttt{D}, and \texttt{F} in Figure~\ref{fig:example}.

Let us examine the first command that \riker skips, the instance of \cmd{cc1} labeled \texttt{A} in Figure~\ref{fig:example}.
We include an excerpt of its \traceir below:

\begin{lstlisting}[language=TraceIR, numbers=left, xleftmargin=0.5cm, escapechar=|]
cc1_1 = Launch(gcc_0, |\Suppressnumber|
  "cc1 x -o /tmp/ccnYMCqc.s", [...]) |\Reactivatenumber|
r71 = PathRef(cc1_1, CWD, "x.c", r--) 
ExpectResult(cc1_1, r71, SUCCESS)
MatchMetadata(cc1_1, r71, |\label{line:cc1-metadata}\Suppressnumber|
 [uid=100, gid=100, type=file, perms=rw-rw-r--]) |\Reactivatenumber|
MatchContent(cc1_1, r71, |\label{line:cc1-content} \Suppressnumber|
 [mtime=1619457130, hash=3c6ea, cached=false]) |\Reactivatenumber|
r75 = PathRef(cc1_1, r3, "tmp/ccnYMCqc.s", |\Suppressnumber|
 -w- truncate create (rw-rw-rw-)) |\label{line:cc1-write1} \Reactivatenumber|
ExpectResult(cc1_1, r75, SUCCESS)
UpdateContent(cc1_1, r75, [hash=054521]) |\label{line:cc1-write2}|
\end{lstlisting}

\riker records the launch of the \cmd{cc1} command and its read of \file{x.c}'s contents and metadata (lines~\ref{line:cc1-metadata}--\ref{line:cc1-content}).
\cmd{cc1} also writes to a temporary file, \file{tmp/ccnYMCqc.s}.
Its first write (line~\ref{line:cc1-write1}) is the result of \cmd{cc1} opening the file and truncating it to zero bytes.
The second write (line~\ref{line:cc1-write2}) emits the generated assembly.

\cmd{cc1} takes two inputs, \file{x.c} and \file{tmp/ccnYMCqc.s}.
The file \file{x.c} is unchanged in the running example.
\file{tmp/ccnYMCqc.s} is different;
it was created by \gcc and is reused by every \cmd{cc1} process started by \gcc.
Because \riker observes that \cmd{cc1} always completely overwrites (truncates) the contents of the file (line~\ref{line:cc1-write1}), \riker concludes that \cmd{cc1} does not depend on the file's earlier state.
As a general rule, whenever \riker can restore output for a command whose inputs do not change, the command can be skipped.
\riker can restore state for any file operation (\sectref{sec:traceir}), so \cmd{cc1 x.c} is skipped.
Similar reasoning allows \riker to skip commands \texttt{C}, \texttt{D}, and \texttt{F} in Figure~\ref{fig:example}.

%\riker can restore an artifact's state by \emph{committing} a saved copy of that artifact to the filesystem.
%Restoring output artifacts is necessary because downstream commands may consume outputs from a skipped command;
%in this case, \cmd{as} depends on \file{tmp/ccnYMCqc.s}.
%Empty files can always be trivially committed.
%File states with actual contents require more work.
%% ---the BLAKE3 hash algorithm~\cite{10.1007/978-3-642-38980-1_8}---
%\riker restores artifact contents using \emph{fingerprinting} and a \emph{caching} mechanisms.
%Fingerprints facilitate fast, accurate detection of file changes, and also provide a straightforward mechanism for saving and restoring artifact states from cache.
%Observe that on line~\ref{line:cc1-write2}, \riker stores the file's hash.
%When evaluating a side-effecting predicate, if the artifact matching the hash is available, then the \traceir step evaluates to true.
%On rebuild, \riker observes that all of the command's predicates evaluate to true, and so \cmd{cc1} can be skipped.

\paragraph{Example 2: Making an inconsequential change.}
Suppose we add a comment to \file{x.c} and run the build again.
This change will have no effect on the final compiled program.
Ideally, an incremental build should exploit this fact to limit work.
Referring again to the previous trace, \riker detects a change (line~\ref{line:dot-content}) for \cmd{cc1} because \file{x.c} changes.
However, \riker correctly halts the build without ever running \cmd{as}, \cmd{collect2}, or \cmd{ld}.
We describe an excerpt of the \cmd{as} command's transcript:

\begin{lstlisting}[language=TraceIR, numbers=left, xleftmargin=0.5cm, escapechar=|]
as_1 = Launch(gcc_0, |\Suppressnumber|
  [as -o /tmp/ccGX6i4a.o], [...]) |\Reactivatenumber|
r26 = PathRef(as_1, r3, "tmp/ccGX6i4a.o", |\Suppressnumber|
  rw- truncate create (rw-rw-rw-)) |\Reactivatenumber|
ExpectResult(as_1, r26, SUCCESS)
r27 = PathRef(as_1, r3,
  "tmp/ccnYMCqc.s", r--) |\Suppressnumber|
ExpectResult(as_1, r27, SUCCESS) |\Reactivatenumber|
MatchMetadata(as_1, r27, [uid=100, |\label{line:as-r27-meta} \Suppressnumber|
  gid=100, type=file, perms=rw-------])|\Reactivatenumber|
MatchContent(as_1, r27, [mtime=1619458806, |\label{line:as-r27-content}  \Suppressnumber|
  hash=10732f, cached=true]) |\Reactivatenumber|
UpdateContent(as_1, r26, [mtime=1619458806,|\label{line:as-r26-update} \Suppressnumber|
  hash=3814e7, cached=true]) |\Reactivatenumber|
\end{lstlisting}  

The \cmd{cc1} command writes output to the file, \file{tmp/ccnYMCqc.s}, which is an input to \cmd{as}.
Since the metadata and content for \file{tmp/ccnYMCqc.s} match their previous values (lines~\ref{line:as-r27-meta}--\ref{line:as-r27-content}), and because \riker can restore the output of \cmd{as} from cache (line~\ref{line:as-r26-update}), \cmd{as} can be skipped.

Encoding this kind of short-circuit behavior in \make is difficult because it exclusively relies on file modification times.
\riker does not have this limitation, only doing work where it matters, in a manner that is completely transparent to the user despite the fact that our \rikerfile still remains unchanged.

\paragraph{System dependencies.}
For clarity the examples in this section omit substantial detail about system dependencies.
The omitted details enable \riker to exploit optimizations well beyond those encoded in a typical \Makefile, while ensuring that no changed dependency is ever missed.

For example, every C build includes system files, like library include files in \file{/usr/include}, and the \cmd{gcc}, \cmd{cc1}, and \cmd{as} files in the compiler toolchain.
%Traces also carefully observe the results of syscalls.
\riker will correctly rerun commands when libraries and header files are updated, or when a new compiler is installed to a location with higher precedence in a user's \texttt{\$PATH}.

\subsection{Summary}
\riker produces fine-grained incremental rebuilds from coarse build specifications, even those comprised of a single command.
Performance does not come at the cost of simplicity.
\riker always ensures that incremental rebuilds produce the same effect as full builds.
Finally, \riker's tracing facility is language-agnostic, giving developers the flexibility to write build scripts in their language of choice.

%In the next section (\sectref{sec:algorithms}), we describe the \riker algorithm in detail.
%Implementation details are discussed in~\sectref{sec:implementation}.
%Finally, we report on \riker's performance in~\sectref{sec:evaluation}.

\section{\traceir}
\label{sec:traceir}

\riker has three design goals:
builds should be easy to specify, always correct, and fast.
%The \traceir language emerged after several abortive attempts to address all three of these concerns using the traditional dependence graph representations.
In this section we describe the basic problem of \emph{forward builds}, how \traceir solves that problem, and how \riker generates \traceir.

\subsection{Forward Builds}

%\traceir was designed to efficiently solve forward builds.
%We first explain what forward builds are, and how they complicate incremental rebuilds.

\emph{Forward builds} were first described by Bill McCloskey, who designed the \memoize build tool~\cite{memoize}.
In contrast to ordinary build tools like \make, which require that users manually enumerate all of a build's dependencies---and is easiest to do ``backward,'' up the dependence chain---forward builds are simpler.
Instead, users write a script that performs a full build.
Discovering dependencies is left to the build tool itself, eliminating a major source of build errors~\cite{10.1145/3428212}.

%In essence, the forward build problem asks how one can perform an efficient incremental rebuild given only an imperative script and an observed execution of a full build.
%
%Forward builds greatly simplify the building of software because users do not need to specify dependencies---only build steps.
%Since a forward build tool is responsible for discovering dependencies, they are never missed, 

\paragraph{Performance woes.} Developers expect build tools to run quickly.
As with traditional build tools, forward build tools take advantage of the fact that most of the time, developers only need to update a fraction of a full build~\cite{doi:10.1002/spe.4380090402}.
When a user makes a code change and runs their build tool, an incremental build does only the minimum amount of work necessary to bring the build up-to-date.
State of the art forward build tools are severely limited, however, only capable of incrementally executing commands literally written in the build script itself~\cite{memoize, fabricate, rattle}.
Since UNIX utilities frequently delegate work to subcommands, many additional optimization opportunities exist.
\riker substantially improves on the state of the art by exploiting fine-grained dependencies among subcommands, producing highly efficient incremental builds.

\paragraph{Example.} Suppose we have the following \rikerfile written in \cmd{bash}:

\begin{lstlisting}[style=rikerfile]
COMMIT_ID=`git rev-parse HEAD`
if [[ `git diff HEAD` ]]; then
  COMMIT_ID+="-dirty"
fi
echo '#define COMMIT_ID "${COMMIT_ID}"' \
  > version.h
gcc -o program code.c
\end{lstlisting}

This build script creates a \file{version.h} file containing the constant \texttt{COMMIT\_ID}, which is used in \file{code.c}.
If the user makes modifications to anything in the repository, the script appends \texttt{-dirty} to the commit ID.
This build script is inspired by a similar example found in the \Makefile for the popular Redis in-memory data store~\cite{redis}.
To correctly handle this build with \make, the programmer must go to great lengths when declaring dependencies for generating \file{version.h}: all source files and some internal state of the \cmd{git} repository.
As a workaround, Redis circumvents \make logic using a shell script to implement custom change detection.

A forward build system always correctly updates \file{version.h} without a workaround because \file{version.h}'s dependencies are automatically discovered.
However, existing forward build tools cannot build this example incrementally.
First, they only model file state---not paths, directories, pipes, symlinks, or sockets.
Second, they do not model subcommands.
Either limitation is sufficient to prevent an incremental build of this example.
By contrast, \riker correctly and incrementally rebuilds \file{program} with this build script.

%Suppose a programmer runs the above build with \riker in a clean \cmd{git} repository.
%If the programmer immediately runs a rebuild without changing any files, \riker correctly determines that no commands need to run.
%\riker ensures that \file{version.h} and \file{program} are present and unchanged from their versions at the end of the full build, restoring them from cache if they have been removed or changed.

When a programmer runs the above build with \riker in a clean \cmd{git} repository, \riker correctly determines that no commands need to run.
If a programmer edits \file{code.c} and runs a rebuild, \riker correctly runs \cmd{git diff} and \cmd{git rev-parse} because they depend on the content of \file{code.c}.
Both \cmd{git} commands write to pipes that are read by \rikerfile, so the top-level build script will run, regenerating \file{version.h}.
The \gcc compiler driver does no useful work, so \riker skips it and directly invokes \cmd{cc1} to compile \file{code.c}, \cmd{as} to assemble it, and finally, \cmd{ld} to link \file{program}.

% DWB: this COULD be compelling, but we're short on space...
%Editing \file{code.c} again will cause \riker to rerun \cmd{git diff}, \cmd{git rev-parse}, and \rikerfile.
%However, because the generated \file{version.h} already includes the \texttt{-dirty} suffix in \texttt{COMMIT\_ID} its content is unchanged.
%\riker is then skips all remaining commands in the build.
%\todo[inline]{This is not correct because it misses code.c's changes.  MUSTFIX.}

% NOTE: Undoing all changes to \file{code.c} would correctly re-generate version.h, recompile code.c, etc. but this example is getting a bit long

\paragraph{Challenges.} An incremental build tool must determine what commands need to run when state changes.
Here we describe the challenges inherent in forward build tools.
\traceir is specifically designed to address these challenges.

\paragraph{Complete dependencies.}
Forward build scripts do not explicitly enumerate any dependencies, so to be safe, forward builds must find \emph{all} of them.
In addition to the obvious ones, the example above contains numerous implicit dependencies.
For example, \cmd{git} commands depend on the directory contents of \file{.git} and the working directory, and the build script itself depends on pipe outputs from those commands.
There are also system dependencies, temporary files, pipes set up by subshells and \gcc, assembly, object files, and executables themselves.
In fact, a number of the build's commands in the example are not present in the script: \cmd{cc1}, \cmd{as}, \cmd{collect2}, and \cmd{ld}.
Finally, this build's behavior depends on file contents, file metadata, and exit codes returned by commands.
%All of these dependencies must be checked during a build.

\paragraph{Temporal relationships.}
Paths alone are insufficient to disambiguate certain kinds of dependencies.
This example, like the one in the introduction (\sectref{sec:intro}), exhibits \emph{temporal} dependencies.
First, several commands create and reuse files with the same name, like temporary files.
Logically, these are not the same file:
commands that use them immediately truncate their contents to empty files before writing to them.
Second, a dependence cycle exists because \cmd{collect2} both reads the output of \cmd{ld} and conditionally launches \cmd{ld} again in order to regenerate those files.
This shows that a dependency used at a later point in \emph{time} is not necessarily the same as an earlier one.
%Both issues can be ``explained away'' by observing that a dependency is state at a given point in \emph{time}.

\paragraph{Conditional behavior.}
Builds can exhibit conditional behavior.
The motivating example (\sectref{sec:overview}) uses shell globbing, conditionally compiling files depending on the contents of a directory.
Command exit codes can also short-circuit builds.
%Such behavior must be encoded for forward builds to work correctly.

\subsection{The \traceir language}

\traceir is a domain-specific, linear representation of build logic.
As forward build tools eliminate the need to provide detailed specifications, \traceir is not intended to be manually inspected.
Instead, its design puts a premium on performance: \traceir is a machine-readable IR streamed from disk.

\riker detects changes by comparing an emulated model of the filesystem against the real filesystem.
The \traceir language specifies how modeled state should be updated, and when and how comparisons should be performed.
%Since \traceir largely concerns itself with maintaining \riker's model and detecting changes, we focus on those aspects in this section.
We describe how \riker uses \traceir to carry out rebuilds in~\sectref{sec:building}.

\subsection{Language Properties}

\paragraph{Design choices.} To address the challenges described earlier, we make the following design choices.
\begin{itemize}
	\item
		\emph{Complete dependencies.}
		All state types, or \emph{artifacts}, that could matter to a build are represented.
		In addition to files, artifacts include directories, pipes, sockets, symlinks, and special files (like \file{/dev/null}).
		All artifacts have both contents and metadata, which are modeled separately.
	\item
		\emph{Temporal relationships.}
		Although builds can be concurrent, \riker always observes build events serially (see~\sectref{sec:parallelism}).
		\riker imposes a total order on \traceir statements.
		Referring to an artifact at a given point in time is unambiguous.
	\item
		\emph{Conditional behavior.}
		A \traceir build transcript represents a single, observed path of execution through a build script.
		It does not explicitly encode conditional behavior.
		A single path is sufficient because the only differences that matter are those that occur between the current state and the previously observed build.
		Whenever a change occurs, the build transcript for the affected command is \emph{discarded} and \emph{regenerated}.
		The updated build transcript represents the path taken by the build script during a rebuild.
		This approach ensures that \riker updates build transcripts in time linear to the number of traced commands.
\end{itemize}

\paragraph{Data types.}
\traceir has a small set of data types.
\IRBool, \IRInt, and \IRString have their usual meanings.
\IRRef represents an artifact at a given point in time.
\IRAccessFlags represents UNIX permissions and file access types (e.g., \texttt{read}, \texttt{write}, etc.).
\IRMetadataState represents UNIX metadata and \IRContentState represents artifact-specific content data.
\IRCmdRef represents a command.

\paragraph{Statements.}
Evaluating a \traceir statement returns an outcome.
Some outcomes represent checks while others are references to artifacts.
Some statements also update \traceir's in-memory state model.

For space reasons, we summarize \traceir statements at a high level according to three logical groupings:
those that record \emph{artifact accesses}, those that \emph{check state}, and those that \emph{update state}.
Every \traceir statement is associated with exactly one executing command.

\begin{itemize}
	\item
		\emph{Artifact access.}
		\riker represents artifact accesses using the \texttt{Ref} family of statements.
		For example, \cmd{PathRef} captures when a command accesses an artifact at a given path from a given directory with the given access flags.
		It returns a reference to an artifact, if it can be resolved.
%		\begin{lstlisting}[language=TraceIRSmall]
%PathRef(cmd:|\IRCmdRef{}|, base:|\IRRef{}|,
%        path:|\IRString{}|, flags:|\IRAccessFlags{}|): |\IRRef{}|
%\end{lstlisting}
		In addition to returning artifact references, evaluating access statements puts artifacts into \riker's model.
		Other accesses include \texttt{FileRef}, \texttt{DirRef}, \texttt{PipeRef}, \texttt{SymlinkRef}, and \texttt{SpecialRef}.
		\riker currently models sockets as pipes.
	\item
		\emph{State checks.}
		The purpose of a state check is to compare modeled state against real state.
		For example, the \texttt{MatchContent} statement checks that the state referred to by the given command matches the state expected by the previous build.
		A content change is always checked first by comparing \texttt{mtime}, and if it is different, by checking a \blake hash value~\cite{10.1007/978-3-642-38980-1_8}.
%		\begin{lstlisting}[language=TraceIRSmall]
%MatchContent(cmd:|\IRCmdRef{}|, ref:|\IRRef{}|, s:|\IRContentState{}|): |\IRBool{}|
%\end{lstlisting}
%		Evaluating a state check returns an \IRBool signaling whether a command would observe a change.
		Other checks include \texttt{CompareRefs}, \texttt{ExpectResult}, \texttt{MatchMetadata}, and \texttt{ExitResult}.
	\item
		\emph{State updates.}
		State update statements record precisely how a command alters system state.
		For example, \texttt{UpdateContent} updates the content referenced by the given command with the given state.
%		\begin{lstlisting}[language=TraceIRSmall]
%UpdateContent(cmd:|\IRCmdRef{}|, ref:|\IRRef{}|, s:|\IRContentState{}|): |\IRBool{}|
%\end{lstlisting}
		Evaluating a state update statement changes \riker's build model.
		Like state checks, they can also signal a change (e.g., if an action fails).
		Other state updates include \texttt{UpdateMetadata}, \texttt{AddDirEntry}, \texttt{RemoveDirEntry}, \texttt{Launch}, \texttt{Join}, \texttt{UsingRef}, \texttt{DoneWithRef}, and \texttt{Exit}.
\end{itemize}

\subsection{Generating \traceir}
\riker generates \traceir whenever a command is executed.
We describe the criteria that \riker uses to execute commands in~\sectref{sec:building}.
\riker gathers the information needed to generate \traceir using a lightweight tracing mechanism that observes the system calls made by an executed command (see~\sectref{sec:implementation}).

During a full build, all commands in the user's \rikerfile are executed and traced.
On rebuild, only the commands that need to run are executed and traced.
Whenever \rkr is invoked, \riker attempts to read in a saved build transcript.
When \riker cannot find a build file, it creates a new, empty transcript.
A full build is simply a degenerate case of a rebuild with an empty build transcript.

Although \traceir is derived from traces of syscalls, an important insight of this work is that most syscall information is irrelevant for the purposes of change detection.
Many syscalls (e.g., \cmd{stat}, \cmd{fstat}, \cmd{lstat}, \cmd{fstatat}, and so on) are variations on the same idea.
Therefore, \traceir is a distillation of syscall information, capturing only the essential dependence information needed to correctly build software.
For example, a typical \cmd{stat} call will generate \IRPathRef, \IRExpectResult, and \IRMatchMetadata statements.

To minimize overhead, \riker streams build transcripts, both while reading them and while writing them.
Although \riker needs space linearly proportional to the number of artifacts in the worst case to store a build model, because of streaming, it needs only constant space to read and write build transcripts.
This optimization is possible because \riker evaluates \traceir sequentially.

%\subsection{Weak Build Equivalence}
%\label{sec:weakequiv}
%
%\subsection{Useful Properties}
%
%
%
%\subsection{Future Work: Program Transformations}
%
%One benefit is that it is easy to write program transformations for \traceir.
%For example, \riker currently uses a simple IR filter to improve performance by removing repeated \traceir statements that result from multiple reads or writes through the same reference.
%This program transformation took an afternoon to write.
%Another benefit is that \traceir is compact.
%A typical build executes millions of syscalls.
%Since \traceir is small, \riker is able to use a concise, packed binary format to keep on-disk build transcript files small.

\section{Build Process}
\label{sec:building}
  
 \begin{figure}[t]
	\small
	\begin{center}
	  \begin{codebox}
		\Procname{$\proc{DoBuild}(trpath)$}
		  \li $i \gets 1$, $M \gets$ \mbox{\{\;\}}, $R \gets$ \mbox{\{ \}}\label{line:incrbuild}
		  \li $T \gets$ \proc{LoadTrace}($trpath$)\label{line:loadtrace}
		  \li \If $|T| \isequal 0$ \kw{then} $T \gets$ \mbox{\{ \IRLaunch(\rkr, \rikerfile, ...) \}}\label{line:cleanbuild}
		  \li \Repeat \label{line:whilecommands}
			\li $M \gets$ \proc{Sync}($M$)\label{line:sync1}
		  	\li $(M, T, D, R) \gets$ \proc{EvalTrace}($M$, $T$, $R$, \mbox{false})\label{line:emulate1}
			\li $R \gets$ \proc{Plan}($D$, $R$)\label{line:plan}
			\li $i \gets i + 1$
		  \li \Until $|R| == 0$\label{line:whileend}
		  \li \proc{CommitAll}($M$)\label{line:commitall}
		  \li \If $i > 1$ \Then\label{line:post-build}
			\li $(\_, T, \_, \_) \gets$ \proc{EvalTrace}($M$, $T$, $R$, \mbox{true})\label{line:emulate2}
		  \End
		  \li \proc{WriteTrace}($T$)\label{line:writetrace}
	  \end{codebox}
	\end{center}
	
	\begin{center}
	  \begin{codebox}
		\Procname{$\proc{EvalTrace}(M, T, R, post)$}
		\li $T' \gets$ \mbox{nil}, $D \gets$ \mbox{\{\;\}}
		\li $R_{pre} \gets \mbox{\{\;\}}, R_{post} \gets \mbox{\{\;\}}$
		\li \For $t$ \kw{in} $T$ \Do
			\li $c \gets \CmdOf{t}$
			\li \If $c \notin R$ \Then \label{line:skip}
				% NOTE: D only contains the commands whose outputs we cannot commit
				\li $(ts, M, D, \delta_{pre}, \delta_{post}) \gets \EvalStmt{t}{M}{R}{D}{post}$ \label{line:eval}
				\li \If $\delta_{pre}$ \kw{then} $R_{pre} \gets R_{pre} \cup \{ c \}$
				\li \If $\delta_{post}$ \kw{then} $R_{post} \gets R_{post} \cup \{ c \}$
				\li $T' \gets T' \mbox{@} \; ts$
			\End
	    \End
	    \li \Return ($M$, $T'$, $D$, $R_{pre} \cap R_{post}$)
	  \end{codebox}
	\end{center}
	
	\caption{
	  \small
	  \riker's build algorithm.
	  \proc{DoBuild} evaluates an entire \traceir build transcript $T$ in the modeled environment $M$ repeatedly until it observes no new changes.
	  \proc{EvalTrace} evaluates a single pass through $T$, returning an updated model and trace, a command dependence graph $D$, and the set of commands that must run, $R$.
	  At the end of the \texttt{while} loop, the build is up-to-date.
	  \label{fig:dobuild}
	}
  \end{figure}
  
\riker's build algorithm compares filesystem state to the build transcript and does work until no changes are observed.
A full build occurs only when no saved build transcript exists, in which case, \riker creates an empty build transcript.
\riker always terminates because the number of times a command can be executed is bounded;
commands are nearly always executed exactly once (see~\sectref{sec:exitcodes}).

\riker's build algorithm, shown in Figure~\ref{fig:dobuild}, operates in ``phases,'' where each phase is an evaluation of an entire \traceir build transcript, $T$, in the context of the model, $M$.
We abstract $T$ here as a list for narrative simplicity, writing an updated trace ($T'$) out at the end of the algorithm (line~\ref{line:writetrace}).
Currently, $T$ is a streaming data structure, but early versions of \riker used the algorithm exactly as written.
Trace evaluation, carried out by \proc{EvalTrace}, is repeated until the set of commands that must run, $R$, is empty (line~\ref{line:emulate1} of \proc{DoBuild}).
$R$ is empty when no changes are observed.

A ``change'' is observed whenever the model $M$ and the actual state of the filesystem disagree.
The specifics of how a change is detected depends on the semantics of the given \traceir statement.
For example, the \IRMatchMetadata statement reports a change when an artifact's metadata does not match the metadata found on disk.

\proc{EvalStmt} (line~\ref{line:eval}) carries out statement evaluation.
It returns one or more \traceir statements, an updated model $M$, a command dependence map $D$ (see~\sectref{sec:planning}), and two flags denoting whether the statement observed a change, either a pre-build or post-build change.
The returned trace statements are used in the next build phase.
When a command is emulated, its trace steps are simply echoed back.
We describe command execution in~\sectref{sec:launching}.

With the exception of the very first and very last phases, which are always emulated, \riker may execute commands during any phase.
Whether a command is launched or emulated depends on whether it was added to the set $R$ in a previous phase (see~\sectref{sec:launching}).
Consequently, many build phases will contain a mix of both command emulation and command execution.
The reason \riker runs a build transcript $T$ repeatedly is so that it never runs a command that need not run.
For example, suppose $T$ contains two commands, $A$ and $B$, and that $B$ consumes one of $A$'s outputs.
If \riker determines that $A$ must run, must $B$ run?
The answer is ``maybe.''
On rebuild, if $A$ produces the same output that it produced previously, then $B$ may be skippable.
However, the determination about $B$ cannot be made until $A$'s effects are observed.

After a transcript has been evaluated, \riker calls the \proc{Plan} method (line~\ref{line:plan}), which may mark extra commands to run.
We describe the conditions that require extra marking in~\sectref{sec:planning}.

Some additional details in Figure~\ref{fig:dobuild} require explanation.

\proc{Sync} (line~\ref{line:sync1} of \proc{DoBuild}) ensures that the state of the model $M$ matches the state of the filesystem (line~\ref{line:sync1}), since some modeled updates will never have been written to disk and therefore do not matter in the subsequent phase.
\proc{CommitAll} (line~\ref{line:commitall} of \proc{DoBuild}) ensures that all emulated state is written to the filesystem at the end of the last phase.

The final call to \proc{EvalTrace} (line~\ref{line:emulate2} of \proc{DoBuild}) does a \emph{post-build check}.
Post-build \traceir statements provide short-circuit logic to speed up rebuilds.
Running \proc{EvalTrace} after the build is finished produces these post-build statements so that \riker can tell when the output of a command is already in its final state.
Specifically, \riker doesn’t need to rerun a command if
(a) its dependencies are unchanged compared to the \emph{previous} phase of the \emph{current} build and its output was cached, or
(b) its dependencies are unchanged compared to the \emph{very last} phase of the \emph{previous} build and its output was cached.
During the post-build check, which is always emulated, \proc{EvalStmt} generates \emph{two} \traceir statements:
(a) an echoed pre-build statement, generated earlier in the build, and
(b) a post-build statement that represents the state present at the very end of the build.

\subsection{Evaluating \IRLaunch and \IRJoin}
\label{sec:launching}

The \proc{EvalTrace} procedure calls \proc{EvalStmt} to evaluate \traceir statements;
due to space limitations, we describe \proc{EvalStmt} at a high level in~\sectref{sec:traceir}.
However, the \IRLaunch and \IRJoin \traceir statements play a special role in build execution and require special explanation.

\riker begins the actual execution of commands during a build when \proc{EvalTrace} evaluates a \IRLaunch statement.
Each \IRLaunch statement is evaluated in a parent command, and launches a child command.
We know that the parent command does not need to run---\proc{EvalTrace} would have skipped over the statement if the parent was in $R$ (line~\ref{line:skip})---but the child may need to run.
If the child must run, \riker will start the command in a new process with system call tracing.
At this point \riker will continue evaluating statements from the input trace while the child executes in the background.

At some point ahead in the trace, the parent command will have a \IRJoin statement to wait for the child to exit.
Evaluating this \IRJoin statement will actually wait for the executing child command to finish.
While \riker waits for the child command it will handle traced system calls.
On each system call stop, \riker generates new \traceir statements to describe the effects and dependencies of the system call, evaluates those statements with \proc{EvalStmt}, and resumes the tracee.
Note that system call stops can come from any background command, not just the child command this \IRJoin statement is waiting for.
Eventually the child command will exit and \riker will return to \proc{EvalTrace}.

It may be surprising that \riker evaluates the \traceir statements collected from an executing child using \proc{EvalStmt}, but this is critically important.
First, evaluating generated \traceir ensures that \riker's model of the filesystem reflects the updates made by executing commands;
later commands that are not running must see the effects of any command that did run earlier in the build.
Second, this evaluation allows \riker to commit changes from the model to the filesystem (e.g. a write to a file from a command that did not run) as executing commands need them (see~\sectref{sec:caching}).
And finally, evaluating the generated \traceir enables command skipping.
We discuss how caching facilitates command skipping in the next section.

%\todo[inline]{Where is the section on skipping!?}

% Note: it may be easy to explain command skipping here: when we evaluate a Launch from a traced parent with an untraced child, just return to the input trace and evaluate until we see an Exit from the child. The system call details aren't terribly important, and the matching could be discussed elsewhere.

\subsection{Caching}
\label{sec:caching}

\riker uses file caching to speed up builds.
Caching also ensures that incremental builds are correct (see~\sectref{sec:correctness}).
%Commands whose output cannot be cached cannot be skipped.
Recall that \riker cannot skip commands whose outputs are uncached.

Again consider the example shown in Figure~\ref{fig:example}.
Suppose that a user deletes \file{tmp1.o}, leaving the build otherwise as-is as shown in the figure.
\riker need not run any commands to bring this build up-to-date: it simply restores \file{tmp1.o}.

However, were \cmd{as} to have a second changed input, \riker could still avoid work, despite the fact that \gcc reused \file{tmp.s} when generating assembly output.
At the time of a rebuild, the version of \file{tmp.s} that \cmd{as} used to produce \file{tmp1.o} is overwritten with assembly generated from \file{z.c}.
Nevertheless, \riker recognizes that the same file is a different dependency at different points in time.
The correct version is restored from cache, and then only \cmd{as} and \cmd{ld} are rerun.

\riker currently caches files, symlinks, and directories.
It does not currently cache pipe, socket, or special files, although we do not foresee any fundamental limitations that prevent us from implementing them.
\riker conservatively skips commands only when their outputs cannot be restored from cache, so pipes, sockets, and special file dependencies are effectively always changed.
Cached files are stored in the \rkrdir directory in the local directory, and are garbage collected when \riker detects that they are no longer reachable.

\subsection{Tempfile Matching}
\label{sec:tempfiles}

Before \riker can skip (i.e., emulate) a command, it matches its dependencies against outputs from other commands that may have been executed and traced.
This introduces a small complication, because executing commands often assign random names to temporary outputs.
Fortunately, the producing command must communicate this name to the consuming command.
Instead of marking consumers as changed, which is always safe but limits incremental work, \riker tries to match command invocations using a \emph{command invocation template}.
This template includes the name of the function and its command-line arguments.
\riker considers any file artifact that appears in \file{/tmp} to be a temporary file.
Additionally, if a child command accesses temporary content in a previous build, then the content for candidate commands must also match.
With this handling, \riker can recognize that a new command invocation is actually the same as a previous invocation modulo the filename.

\subsection{Build Planning}
\label{sec:planning}

The purpose of build planning (line~\ref{line:plan} in \proc{DoBuild} of Figure~\ref{fig:dobuild}) is to identify additional commands that must run.
These additional commands do not observe changes directly, so they are not marked in \proc{EvalTrace}.
\riker marks these commands for two reasons: to preserve correctness (e.g., to ensure build termination) or to improve efficiency.

\proc{Plan} works much like the mark-sweep garbage collection algorithm~\cite{10.1145/367177.367199} and uses the command dependence graph, $D$, returned by the \proc{EvalTrace} command.
$D$ is a directed graph of producer-consumer relationships between commands.

Commands are marked in the following conditions:
\begin{itemize}
	\item a command consumes uncached input produced by a command already marked to run; % rule 5
	\item a command produces uncached output consumed by another command already marked to run; or % rule 3
	\item a command produces uncached output that should persist at the end of the build (as indicated by a post-build statement).  % rule 2
\end{itemize}

Normally, the above criteria identify commands that subsequent build phases would eventually identify, so marking reduces the number of build phases.
However, without special handling, one dependence structure prevents \riker's algorithm from terminating: \emph{cycles}.
In $D$, a dependence cycle appears as a strongly-connected component (SCC)~\cite{DBLP:journals/siamcomp/Tarjan72}.
Marking causes commands in a dependence cycle to run atomically.

To illustrate, suppose \cmd{A} and \cmd{B} are a SCC.
Initially, neither \cmd{A} nor \cmd{B} are marked to run, but on the first iteration, \cmd{A} observes an input change.
In the second iteration, \riker traces \cmd{A}, emulating \cmd{B} and detecting a change for \cmd{B}.
\cmd{B} is marked to run.
In the third iteration, \riker traces \cmd{B}, emulating \cmd{A} and detecting a change for \cmd{B}.
\cmd{A} is marked to run.
Without corrective action, the build algorithm will loop forever.

\riker needs only atomically run SCCs with uncacheable dependencies; it is not used for file or directory dependencies.

\subsection{Exit Code Handling}
\label{sec:exitcodes}

\riker always executes commands once except under a rare condition involving exit codes.
Programs can act on this information and so exit codes must be viewed as a kind of dependence.
To our knowledge, \riker is the only forward build system that models exit codes correctly.

Because such behavior is rare, \riker speculates, optimistically assuming that a parent command does not depend on a child's exit code.
When the child's exit does not change, then speculation saves work.
When the child's exit code changes, \riker backtracks and executes the parent again.
Executing the parent may re-execute the child if the child's dependencies also change.
In the worst case, \riker could backtrack on every command, taking $O(n^2)$ time, where $n$ is the number of commands.
We have never observed this.
Even for builds that contain compilation errors---and thus changed exit codes---we still observe that the total work done is close to $O(n)$.

\subsection{Correctness}
\label{sec:correctness}

Program correctness is defined as ``whether [a program] accomplishes the intentions of its user.''~\cite{10.1145/363235.363259}.
%Unfortunately, unambiguously inferring the intent of a user is a difficult task, if not impossible in many cases.
With respect to build tools, a user never expects that full and incremental builds produce different outcomes.
Avoiding different outcomes is an explicit goal of the \make tool~\cite{doi:10.1002/spe.4380090402}.
%``Forgetting to compile a routine that has been changed or that uses changed declarations will result in a program that will not work, and a bug that can be very hard to track down.''~\cite{doi:10.1002/spe.4380090402}
When a build algorithm always produces equivalent outcomes for full and incremental builds, we call that algorithm \emph{consistent}.

A build tool that produces inconsistent outcomes is clearly \emph{incorrect}.
Therefore, a practicable definition of correctness for a build algorithm is whether it is consistent.
Since running a command is by definition consistent, whether a build algorithm is consistent hinges on how it handles skipped commands.

\make is not consistent because it can skip a command whose inputs have changed.
Such skips occur whenever a \Makefile misses any dependency, a common mistake~\cite{10.1145/3428212}.
This holds for any build system that permits the same flaw.

By contrast, \riker is consistent.
\riker skips commands with unchanged inputs whose outputs can be restored from cache.
Outputs restored from cache for deterministic commands are trivially equivalent to outputs cached in a previous build because they are identical.
Because some previous build ran the command whose output was cached, deterministic commands are consistent.
Nondeterministic commands produce different outputs for the same input.
Such outputs form an equivalence class we call \emph{weakly equivalent}.
\emph{Any} output from a weakly equivalent set may be returned, since by definition the command \emph{could} return it.
Returning cached outputs ensures consistency for nondeterministic commands.

\section{Implementation Details}
\label{sec:implementation}

% In this section, we describe the important elements of \riker's implementation.
\riker is written in C++17, and includes three third party libraries: \cereal handles serialization, \blake handles hashing, and \cliEleven processes command line arguments~\cite{cereal, 10.1007/978-3-642-38980-1_8, cli11}.
The majority of \riker's algorithms are platform-agnostic, however \riker uses AMD64 Linux-specific tracing facilities.
This section describes command tracing and processing mechanisms in more detail.

\subsection{Tracing Command Execution}
When a command is encountered during a build that needs to be launched, \riker launches it in a tracing environment.
Tracing produces \traceir statements, replacing the current \traceir program for the next build.
Tracing is ``always on'' during execution so that dynamic dependencies are never missed.

\riker intercepts a command's syscalls using both a custom \file{libc} wrapper and \ptrace.
The former is a high-performance userspace mechanism employed for the most common syscalls.
The latter is simpler to implement, but incurs context switch overhead, so is reserved for less common syscalls to ensure completeness.
Some system calls---\fun{getpid} for example---do not depend on filesystem state or interact with other processes, and do not need to be intercepted.
We utilize \seccomp BPF filters to provide a high-performance mechanism for trapping the syscalls that matter---75 functions in our implementation.
Some of the traced system calls are \fun{open}, \fun{close}, \fun{stat}, \fun{read}, \fun{write}, and their many variations; those that create pipes, links, and directories; and those like \fun{fork}, \fun{exit}, \fun{wait}, and \fun{exec}.

\subsection{Lightweight \file{libc} Wrapper}
To reduce the overhead of system call tracing with \fun{ptrace}, \riker injects a small shared library into the commands it launches.
This mechanism is inspired by \rr, which combines wrappers with binary rewriting to intercept system calls without \ptrace stops~\cite{203227}.

The shared library intercepts calls to a small number of \file{libc} wrappers around system calls.
On system call entry, the library sends a notification to \riker using a shared memory channel.
After \riker processes the system call entry, the shared library issues the actual system call from a fixed address excluded from tracing by \riker's \seccomp BPF program.
During development we found that this alternative approach reduces tracing overhead by at least \wrapperOverheadReduction.

\subsection{Parallelism}
\label{sec:parallelism}
Tracing is single-threaded to ensure that transcripts capture a canonical ordering of system calls.
When two commands write concurrently, \riker records those writes in the observed order~\cite{10.1145/3335772.3335934}.
\riker only suspends traced commands when they perform system calls.
This has the effect of serializing system calls, but it does not prevent parallel execution.

After launching a traced process, \riker continues to emulate other \traceir statements.
\riker processes trapped system calls whenever the emulated command that launched them encounters an \IRJoin step.
During a stop, a syscall-specific handler generates the appropriate \traceir.
Consequently, any \rikerfile that launches commands in parallel still runs in parallel between system calls.
Since build tools spend most of their runtime in userspace, and emulation is fast, this design imposes little overhead.

\section{Evaluation}
\label{sec:evaluation}

Our evaluation of \riker addresses four key questions:

\begin{description}
  \item[RQ1:] Are \riker builds easy to specify?
  \item[RQ2:] Are full builds with \riker fast enough?
  \item[RQ3:] Does \riker perform efficient incremental rebuilds?
  \item[RQ4:] Are \riker builds correct?
\end{description}
  
To answer each of these questions, we use \riker to build \numbench software packages, including large projects like LLVM, memcached, redis, and protobuf.
Evaluation was conducted on a typical developer workstation with an Intel Core i5-7600 processor, 8GB of RAM, and an SSD running Ubuntu 20.04 with kernel version 5.4.0-80.
Builds use either \texttt{gcc} version 9.3.0 or \texttt{clang} 10.0.0;
other tools that run during the build are the latest versions available in standard Ubuntu packages.

%%%%%%%%%%%%

\subsection{Are \riker builds easy to specify?}
\label{sec:eval-easy}

To answer this question, we wrote \rikerfile{}s for seven applications: lua, memcached, redis, \rkr, sqlite, vim, and xz.
The new builds produce the same targets as the projects' existing \make or \cmake builds.
Unlike the default build systems, the \riker-based builds do not list any dependencies or incremental build steps.
Three of these builds were written by undergraduate students over the course of a few days;
the students were new to \riker and unfamiliar with the project sources they were building.
The biggest challenge the students faced was understanding the existing build specifications, a task that is likely easier for the project's own developers.

A key simplifying feature of a \rikerfile is its brevity, making it easier to understand and check.
The \Makefile for memcached is just under 151KB long, and is difficult to understand.
Here, we include memcached's \emph{entire} \rikerfile:

\begin{lstlisting}[style=rikerfile,escapechar=§]
CFLAGS="..."
DEBUG_CFLAGS="..."
MEMCACHED_SRC="memcached.c hash.c ..."
TESTAPP_SRC="testapp.c util.c ..."

gcc $CFLAGS -o memcached $MEMCACHED_SRC -levent
gcc $DEBUG_CFLAGS -o memcached-debug $MEMCACHED_SRC -levent
gcc $CFLAGS -o sizes sizes.c -levent
gcc $CFLAGS -o testapp $TESTAPP_SRC -levent
gcc $CFLAGS -o timedrun timedrun.c -levent
\end{lstlisting}

This level of simplification is typical.
Our largest \rikerfile---used to build sqlite---is just over 5KB compared to the original 46KB \Makefile.
These reductions, combined with our experience writing {\rikerfile}s for large projects, is strong evidence that \riker builds are easy to specify.

%%%%%%%%%%%%

\subsection{Are full builds with \riker fast enough?}

The first full build of any software project is likely to be the longest build.
Full builds are also where \riker incurs the largest overhead in real seconds.
Importantly, full builds are not the common case.
Developers run incremental builds far more often than full builds.
This section shows that even with the extra delay, full builds are acceptably short.
%---but full builds with \riker have to complete in a reasonable amount of time.

To measure \riker's overhead, we built \numbench software projects with \riker.
Seven of these projects use a \rikerfile that replaces the default build (see~\sectref{sec:eval-easy}), while the other half use a \rikerfile that \emph{wraps} the default build.
The only requirement for a \rikerfile is that it run a full build;
we can trivially do a full build for a \make project with the following \rikerfile:

\begin{lstlisting}[style=rikerfile,escapechar=§]
make --always-make
\end{lstlisting}

\riker is still discovers incremental build steps and dependencies with this \rikerfile, although it is likely to incur additional overhead because \make itself is running under \riker's tracing.

\begin{figure}[t]
  \begin{center}
    %% overhead plot
	\iftoggle{arxiv}{
	  \includegraphics{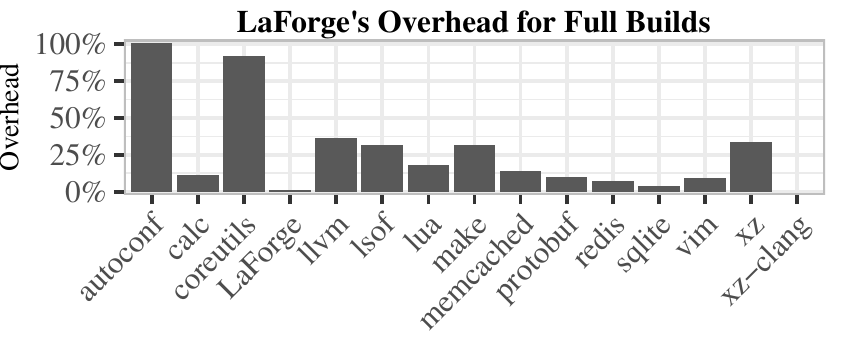}
	}{
	  \includegraphics{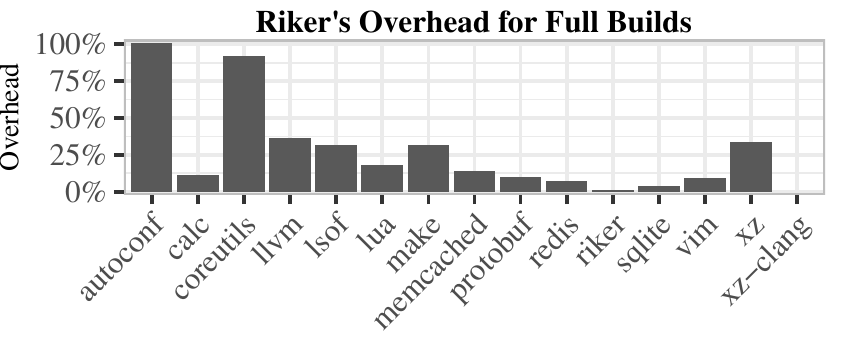}
	}
    \caption{
      \riker runtime overhead for full builds compared to each project's default build system.
      Median overhead is \percentslowdown.
      For all but three benchmarks, full builds took $<6s$ longer.
      coreutils, protobuf, and LLVM took 38s, 58s, and 6 minutes longer, respectively.
      \label{fig:overhead}
    }
  \end{center}
\end{figure}

Figure~\ref{fig:overhead} shows the results of running full builds with \riker.
Each project is built five times with \riker and its default build system.
The overhead is the median \riker build time divided by the median default build time minus one.
Median full-build overhead for all benchmarks is just \percentslowdown;
most builds have between 7\% and 34\% overhead.
\traceir transcript sizes are roughly proportional to build time, ranging from 2MB for autoconf (1.2s build) to 264MB for LLVM (25 minute build).
In absolute terms, \riker spends a median of just 1.5 seconds longer to perform a full build than each project's default build system.
The longest additional waiting time for a \riker build is for LLVM, which takes about six minutes longer than the default 25 minute build (34\% overhead).

Building \riker has the lowest overhead while autoconf and coreutils have the highest.
The worst overheads are when tracing \cmd{gcc}, which issues an order of magnitude more system calls than \cmd{clang}.
The difference is even larger for short compilations.
\riker is built with \cmd{clang}, while coreutils and autoconf both use \cmd{gcc}.
This effect is largest with xz, which uses \cmd{gcc} by default.
Replacing \cmd{gcc} with \cmd{clang} when building xz significantly reduces \riker's overhead.
A \cmd{clang} build of xz with \riker is actually \emph{faster} than the default build using \cmd{gcc}.
Given that compiler choice appears to have a larger effect on build time than using \riker, and many developers still choose to use \cmd{gcc} rather than \cmd{clang}, these overheads are well within the range of acceptable overheads for most projects.

%%%%%%%%%%%%

\subsection{Does \riker perform efficient incremental rebuilds?}

The most important measure of efficiency for a build system is its ability to perform fast incremental rebuilds.
We perform two experiments to measure the efficiency of \riker's incremental builds.

First, we measure the time it takes \riker to perform a \emph{no-op build}---one where no commands need to run---by running an incremental build immediately after finishing a full build.
\riker must run the entire build algorithm to confirm that no commands need to run.
The median \riker no-op build time over the \numbench benchmarks is just 218ms, compared to 5ms for the default build.
The longest additional wait is for the LLVM build, which takes 12.8s with \riker compared to 4.8s with \make.
Most no-op builds with \riker take just 162ms longer than the default build system, an imperceptible difference.

Second, we use real developer commits to measure the efficiency of performing incremental builds with \riker versus a project's default build.
We run this experiment on six projects---memcached, redis, \rkr, sqlite, vim, and xz---all of which have custom \rikerfile{}s.
We perform a full build of each project, and then measure the time and number of commands required to update the build over the next 100 commits in the project's \cmd{git} repository.
This experiment simulates a developer performing incremental builds after editing a subset of the project's source files.

\begin{figure}[t]
  \begin{center}
    %% overhead plot
	\iftoggle{arxiv}{
	  \includegraphics{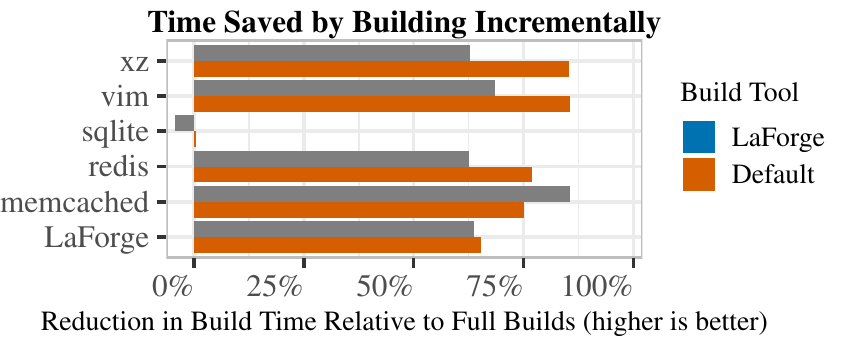}
	}{
	  \includegraphics{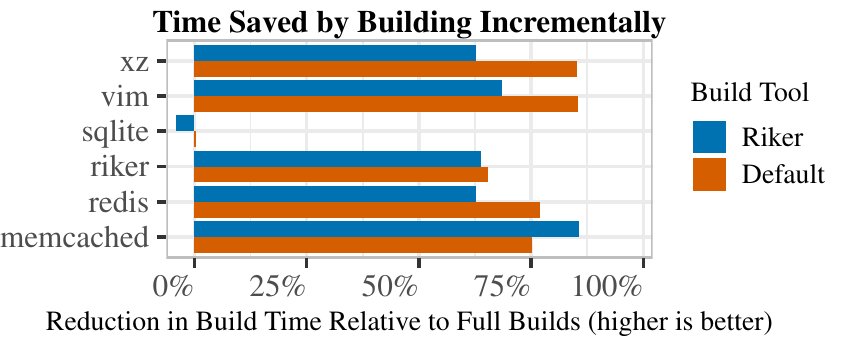}
	}
    \caption{
      Build time saved by running incremental builds over 100 commits of each project.
      Time saved is the percent reduction in build time compared to running a full build at each commit using the project's default build system.
      Most incremental builds with \riker take less than \absslowdown longer than the default build.
      \label{fig:savings}
    }
  \end{center}
\end{figure}

Figure~\ref{fig:savings} shows the results of this second experiment.
The graph shows how much time incremental builds save compared to running a full build at each commit (higher is better).
Note that we compare \riker's incremental build times to the time it would take to run a full build with the project's \emph{default} build system.
This ensures \riker's overhead on the full build does not give it an advantage compared to the slightly faster default build system.
In absolute terms, \riker completes 95\% of its incremental builds within 5s of the default build system;
one outlier in the vim project \emph{may} be a case where the default build is missing a dependency, but we have not confirmed this.

In every benchmark except one (sqlite), \riker is able to save at least 60\% of full build time.
Over these five benchmarks (excluding sqlite), incremental builds with the default build system save 77.6\% of build time, compared to 68.7\% for \riker.
This amounts to a total savings of roughly four hours and five minutes for the default build system, versus three hours and 37 minutes for \riker.
We have no way of measuring time spent troubleshooting build systems, but it seems plausible that developers on these five projects spent at least 28 minutes working on their complex build systems over these 100 commits.

The sqlite benchmark is an interesting case where neither \riker nor the default build system saves any work.
This is because sqlite's build concatenates all of its source files together before compiling them.
No incremental compilation is possible.
\riker ``saves'' -4\% over the full sequence of 100 commits for sqlite, but this slowdown amounts to less than three additional seconds for a rebuild.

Considering \riker's build specifications include \emph{no explicit incremental build steps}, these results are extremely impressive.
Every bit of work that \riker saves is determined \emph{automatically} by its build algorithm.

%%%%%%%%%%%%

\subsection{Are \riker builds correct?}

We run each project's full test suite for both original and \rikerfile builds.
For the six projects in the previous section, the tested outputs are the product of one full build and 100 incremental builds, one for each commit.
The remaining projects run only full builds.
Every \riker project passes exactly the same tests as the original build system.

%%%%%%%%%%%%

\subsection{Evaluation Summary}
Our evaluation shows that when using \riker, the resulting builds are simpler, nearly as fast, and always correct.
%With just a few lines, \riker produces incremental builds comparable to those of much more complicated specifications.
\riker's benefits far outweigh its modest overheads, particularly given that merely choosing \gcc over \clang has more of a performance impact than choosing \riker over \make.

\section{Related Work}
\label{sec:related}

\make is one of the earliest and most widely-deployed build automation tools~\cite{doi:10.1002/spe.4380090402, RMCH}.
\make takes a file called a Makefile as input, which explicitly encodes the relationships between build outputs and their dependencies.
%A \Makefile is written in a domain-specific superset of the UNIX shell language.
%As large software projects are often composed in a layered fashion, \make projects can be defined recursively, although doing correctly is tricky and is sometimes discouraged~\cite{RMCH}.
One of the most important features of \make is its ability to produce incremental builds.
%in other words, \make rebuilds only the necessary parts of a build by tracking which dependencies have changed.
Because numerous similar build systems exist, we focus on the most notable alternatives.

Several build systems address build performance.
\tup introduces a build language that improves change detection for faster build times~\cite{tup}.
\shake lets users write builds using arbitrary Haskell functions~\cite{Mitchell:2012:SBB:2364527.2364538}.
\shake is faster than \make in some cases, and constructs the build's dependence graph dynamically, allowing generated files to be encoded as dependencies.
% This feature is useful in cases where dependencies are hard to specify ahead of time, such as platform-dependent header files.
% \shake is faster than \make for large, recursively-defined software builds.
\pluto surfaces more granular dependencies through an automatic dynamic dependency analysis that enables better incremental builds~\cite{Erdweg:2015:SOI:2814270.2814316}.
Unfortunately, \pluto is not language-agnostic, and requires language-specific extensions to support new languages.
Finally, the \ninja build system is a low-level build language that focuses on speed and is intended to be generated by high-level build tools~\cite{ninja}.
Unlike \riker, all of these systems require manual dependency specification.

Several tools focus on providing builds as a cloud service.
One of the earliest is the \vesta system, which includes a filesystem at its core and can perform continuous integration-like tasks, including building software, using a specialized modeling language~\cite{10.5555/993815}.
More recently, \buck, \bazel, and \cloudbuild offer performance improvements by hosting builds on high-performance clusters~\cite{buck,bazel,10.1145/2889160.2889222}.
\cloudbuild focuses on making it easy to import existing build specifications from software projects and provides a set of heuristics that attempt to infer commonly missing dependencies.
\buck and \bazel notably feature bit-for-bit reproducible builds.
A \emph{reproducible build system} compiles code deterministically, ensuring that a given input on the same architecture will always produce the same binary.
% Importantly, reproducible build systems facilitate debugging since errors detected in production environments but not found during development can be confidently attributed to differences in environment.
%Reproducible build systems facilitate debugging and also help mitigate software vulnerabilities that target build systems themselves, such as recently-demonstrated \emph{supply-chain attacks} or Ken Thompson's infamous compiler hack~\cite{reed2014supply, Thompson:1984:RTT:358198.358210}.
All these tools require that users manually specify dependencies, unlike \riker.
% Users of \buck and \bazel must specify dependencies manually, although \buck uses a set of annotations that gives the tool's incremental update algorithm access to more granular dependence information than is explicitly specified.
%The emphasis of the above work is on build services, and unlike \riker, they require that users manually specify dependencies.

\emph{Forward build systems} like \rattle, \memoize, and \fabricate completely free users from specifying dependencies; instead users provide a sequence of build commands~\cite{rattle, memoize, fabricate}.
\rattle's specifications are written in a Haskell DSL, and it uses a form of speculative execution to improve build performance.
\memoize and \fabricate are language-agnostic, allowing any executable program to function as a specification.
The common theme of these tools is that they automatically discover dependencies using tracing facilities.
\fabricate extends \memoize to Windows, and also provides some facilities for specifying parallel builds.
Unlike \riker, these tools cannot automatically discover fine-grained incremental builds;
\rattle, \memoize, and \fabricate can only select from commands explicitly provided by users.
Furthermore, none of these projects model all filesystem state, which means that they can fail to update projects.

% we say all of this earlier
%\riker is a forward build system that improves on prior work in three notable respects.
%First, \riker is easy to use by being language-agnostic.
%Users can write build scripts in their language of choice.
%Second, \riker produces incremental builds from very coarse build specifications while capturing \emph{all} dependencies.
%Finally, \riker is the only build system that correctly produces fine-grained incremental builds.

\section{Conclusion}
\label{sec:conclusion}

\riker significantly lowers the burden of building software projects while providing always-correct, high-performance builds for free.
Migrating to \riker from an existing build system is easy, and we plan to extend \riker in the future to make migration even simpler.
As incremental builds are for the developer's benefit, and a \rikerfile{}s is a standalone full build script, developers need not deploy \riker to end users.

\section{Acknowledgments}
\label{sec:conclusion}

This work was supported by National Science Foundation grant CNS-2008487.
Alyssa Johnson and Jonathan Sadun helped produce an early prototype of this work.
We also thank John Vilk, Benjamin Zorn, and Emery Berger for their proofreading assistance.

\bibliographystyle{plain}
\bibliography{references}

\end{document}